\documentclass[epj, nopacs]{svjour}
\usepackage{graphics}
\usepackage[numbers]{natbib}
\usepackage{amsmath}
\usepackage{hyperref}
\usepackage{breqn}
\usepackage{bm}
\usepackage{float}
\usepackage{amssymb}
\usepackage{dcolumn}

\usepackage{subcaption}

\hypersetup{
    colorlinks=true,
    linkcolor=blue,
    filecolor=magenta,      
    urlcolor=cyan,
    pdftitle={Overleaf Example},
    pdfpagemode=FullScreen,
}

\makeatletter
\g@addto@macro\@floatboxreset\centering
\makeatother

\begin{document}
\title{Potential for Tensor Polarized Deuterons in Hall D at Jefferson Lab}
\titlerunning{Photon Scattering from Tensor Polarized Deuterium}
\authorrunning{Dalton, Deur and Keith}

\author{Mark M. Dalton\inst{1} 
\and Alexandre Deur\inst{1} 
\and Chris Keith\inst{1}
}                     
%
%
\institute{Thomas Jefferson National Accelerator Facility, Newport News, VA 23606, USA 
}
\date{Received: date / Revised version: date}

\abstract{Hall D at Jefferson lab is an ideal place to install a polarized deuteron target which can be ``tensor polarized", allowing the separation of the spin states $m=0,\pm1$ or the measurement of tensor asymmetries.
The bremsstrahlung photon beam with 12 GeV endpoint provides very little heating or radiation damage compared to an electron beam, allowing the target to be run in frozen spin mode.  
Adiabatic fast passage spin manipulations can then be used to greatly enhance the population of the $m=0$ spin state of the deuteron.
Coherent photoproduction of $\rho$ mesons from deuterium is sensitive to double-scattering at high momentum transfer and in the $m=0$ spin state an additional sensitivity at intermediate momentum transfer opens up.
We propose a frozen spin target for Hall D and the measurement of $\rho$ photoproduced coherently from the deuteron as a flagship measurement.
}

\maketitle
\section{Introduction}
\label{sec:intro}

The nucleus can serve as a laboratory to study the space-time development of strongly interacting systems which are not accessible in scattering off nucleons~\cite{Raufeisen:2003sd}.
This is achieved by using the nucleus itself to interact with systems that persist or evolve on scales comparable to the nuclear size.
These ideas have been used to study the coherence length of virtual photons~\cite{HERMES:1998ajz},  the evolution of small sized hadronic configurations involved in color transparency~\cite{HERMES:2002tmh}, and shadowing in low$-x$ deep inelastic scattering~\cite{NewMuon:1996yuf}.
The coherence length of photons and the color transparency have similar dependencies on energy and momentum transfer and therefore, in order to characterize color transparency effects, coherence length effects must be understood.  
Vector meson production with medium energy real photons offers a rigorous test of our understanding of coherence length effects.

Deuterium is an especially important target, as it is the simplest bound state of nucleons.  The wavefunction is well known up to internal momenta well beyond the Fermi momentum.
There are only two nucleons available for interaction.  After production on the first nucleon, a created hadron can only interact with the other nucleon in a ``double-scattering" event.
The spatial relationship between nucleons can be influenced by the nuclear polarization state~\cite{Cosyn:2020kwu}.
In comparison to heavy nuclei, the deuteron has two major advantages.
The charge radius is small enough that the coherence length in $\rho$ meson photoproduction, 
$l_c = 2 \nu / m^2_V$,  
will exceed the deuteron size at $\nu \approx 3$~GeV.  
For the $^{208}$Pb nucleus this is at $\nu \approx 8$~GeV, making studies of the deuteron more tractable at existing facilities.
Also, the second scattering, the signal to be studied, is not complicated by interaction with multiple nucleons and therefore, devoid of those additional theoretical uncertainties.
Thus, the deuteron is an ideal system for studying double-scattering.
Its magnitude contains information about the intermediate hadronic state and its evolution from a compact, color-singlet quark-gluon wave packet to a soft vector meson.
Similar ideas are proposed in electron scattering with the deuteron as a target, to study small size configurations of the proton using color transparency~\cite{Li:2023adv}, essentially by knocking one nucleon into the other.

The CEBAF Accelerator at Jefferson Lab produces a continuous beam of electrons of up to 12 GeV~\cite{Adderley:2024czm}.  
Here, we focus on Hall D, which uses a coherent bremsstrahlung photon beam of up to 12 GeV and contains the GlueX detector~\cite{GlueX:2020idb}---a hermetic spectrometer with good acceptance for both charged and neutral particles.
These capabilities are an excellent match for the physics of interest in tensor-polarized deuterium.

Tensor polarization, denoted here as $Q$, is one of two orientation parameters that are necessary to describe an ensemble of identical spin-1 particles in a magnetic field with cylindrical symmetry.  The other parameter is the more familiar
vector polarization, $P$.  These are defined as: 
\begin{eqnarray}
    P &=& N_{+} - N_{-}  \label{eq:Pz} \\
     Q&=& (N_+ -N_0) - (N_0 - N_-) = 1 - 3N_0 \label{eq:Pzz}
\end{eqnarray}
Here $N_{+}$, $N_{-}$ and $N_{0}$ are the fraction of spins in the $m = +1,-1,0$ projections respectively, normalized such that
\begin{equation}
\label{eq:norm}
    N_{+} + N_{-} + N_{0} = 1.
\end{equation}
Written in terms of the polarizations, the fractional populations are:
\begin{eqnarray}
\label{eq:N+}
    N_{+} &=& \frac{1}{3} + \frac{P}{2} +\frac{Q}{6}, \\
\label{eq:N0}
    N_{0} &=& \frac{1}{3}(1 - Q), \\
\label{eq:N-}
    N_{-} &=& \frac{1}{3} - \frac{P}{2} +\frac{Q}{6}.
\end{eqnarray}
The tensor polarization can thus be viewed as a measure of the $N_{0}$ population.  $Q>0$ indicates a relative depletion of the $m = 0$ projection, while $Q<0$ indicates its enrichment.

In Sect.~\ref{sec:double} we describe a kinematic window for double-scattering that is available when deuterons, polarized parallel to the photon momentum ({\it i.e.}, longitudinal for real photons), are populated in the $m=0$ magnetic substate.  
Experimentally, this implies a polarized deuteron target with negative tensor polarization, a condition that cannot be achieved utilizing standard methods for polarized solid targets.  
However, it can be produced on a frozen-spin target using an NMR technique called Adiabatic Fast Passage (AFP)~\cite{abragam1961principles}.  
The lifetime of this state at the relevant temperature and magnetic field is not yet known, though it is understood to be highly dependent on temperature and should persist long enough to make impactful double-scattering measurements when the sample is cooled to ultralow, frozen-spin temperatures.
Section~\ref{sec:LineShape} describes how the target polarization information is encoded in the NMR lineshape of solid samples.  The production of negative tensor polarization using AFP is detailed in Sect.~\ref{sec:AFP}, and an estimate of the achievable results using a simple model is given in Sect.~\ref{sec:Model}.  Finally, we provide a brief conceptual design of a new polarized target for Hall D at Jefferson Lab that can be used for the proposed double-scattering measurements.

\section{Double-scattering in the Deuteron}
\label{sec:double}

Vector meson photoproduction from polarized deuterium is a good laboratory to study double-scattering~\cite{Frankfurt:1997ss,Frankfurt:1998vx}, where a photon interacts with one of the nucleons to produce an intermediate hadronic state which re-scatters from the second nucleon before forming a final state vector meson.  

\begin{figure}[htb]
\includegraphics[width=0.40\textwidth]{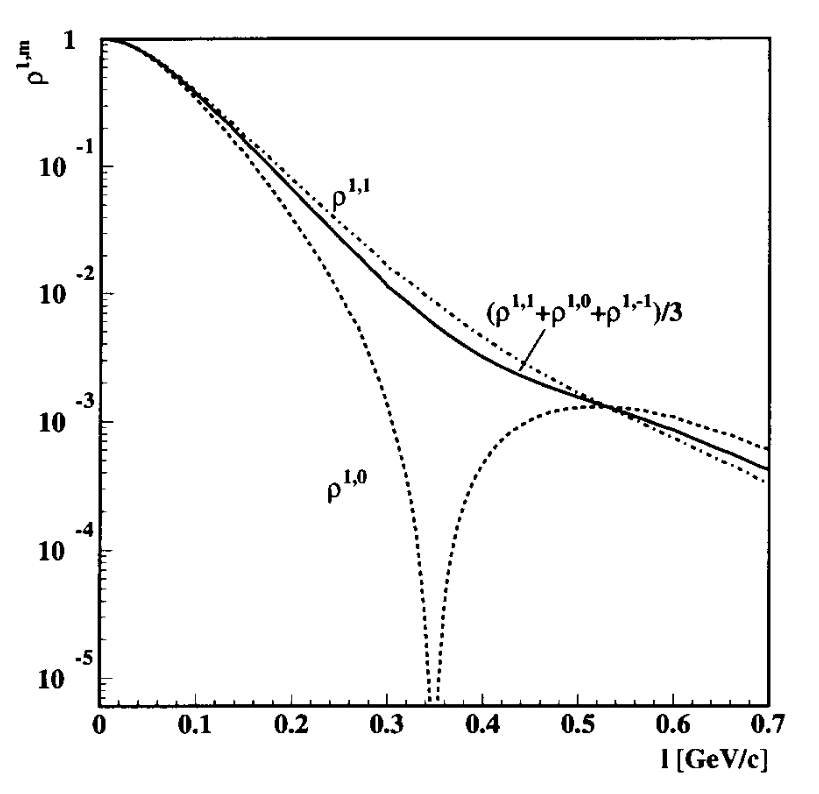}
\caption{The deuteron density matrix, $\rho^{l,m}$, for a polarized deuteron with spin quantization axis parallel to the photon momentum (longitudinal for real photons) versus  $l=|l_{\perp}|/2$, half the magnitude of 3-momentum transfer in the perpendicular direction.   $\rho^{1,0}$ and $\rho^{1,1}$ are for spin projections $m=0$ and $m=1$  (or $Q=-2$ and $P = +1$) respectively.  The solid line is the unpolarized case.  For $m=0$, $\rho^{1,0}$ vanishes at $l_{\perp}\approx0.7$~GeV/c.  Figure from Ref.~\cite{Frankfurt:1997ss}.}
\label{fig:deuterondensity}
\end{figure}

In the polarized case, the Born cross section is proportional to the deuteron density matrix shown in Fig.~\ref{fig:deuterondensity}.  Here it can be seen that the $m=0$ projection has a very strong diffractive minimum, at $l=|l_{\perp}|/2\approx0.35$~GeV, where $l$ is the magnitude of per-nucleon 3-momentum transfer in the perpendicular direction.
This is a relatively small momentum transfer compared to the region where double-scattering dominates for unpolarized $\rho$ photoproduction from the proton ($l_{\perp}^2\approx -t\gtrsim 0.7$\,GeV$^2$)~\cite{Anderson:1971ar}, where $-t$ is the squared four-momentum transfer.
This dip is a kinematic window through which one can study double-scattering.{

\begin{figure}[htb]
	\includegraphics[width=0.50\textwidth]{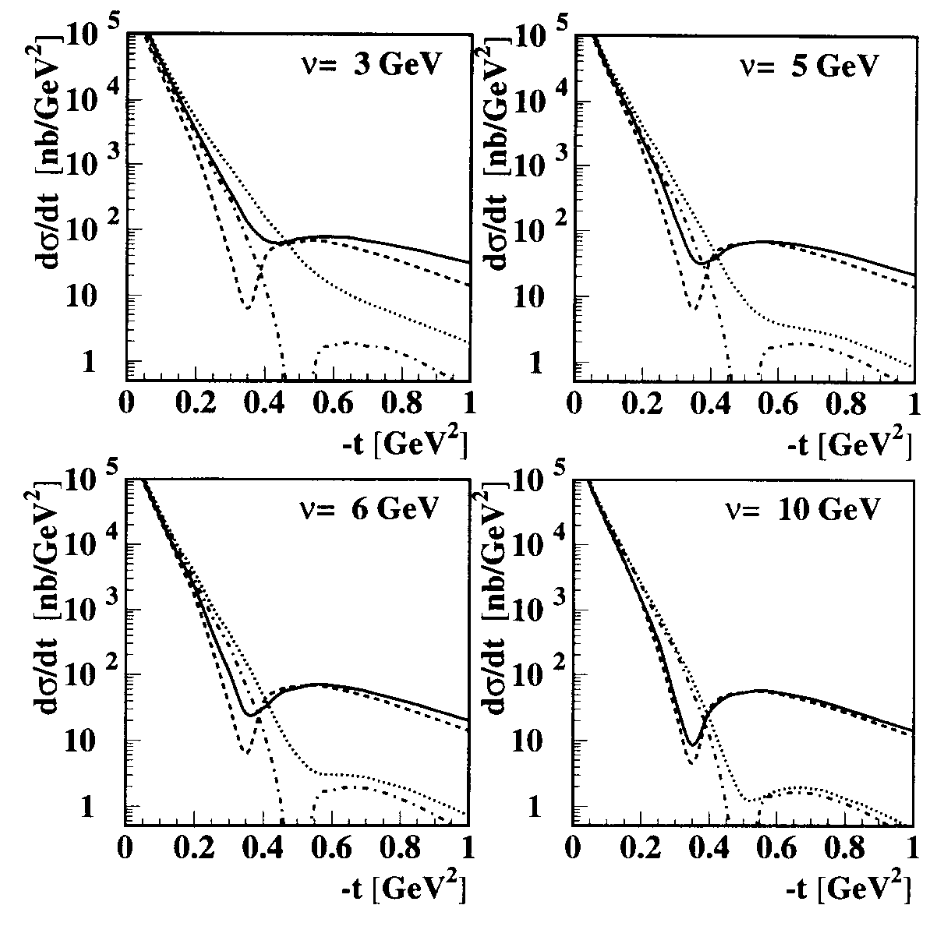}
	\caption{Differential cross section $d\sigma/dt$ for $\gamma \vec{d}\to \rho d$ as a function of the squared four-momentum transfer $-t$, with the deuteron initially in $m=0$ state longitudinally.
    The four panels correspond to four different incident photon energies $\nu$. Solid line: complete vector meson dominance calculation (single and double scattering), dashed line: complete calculation for infinite interaction length (maximum shadowing),
    dotted line: Born contribution (single scattering only), and dot-dashed curve: Born contribution for infinite interaction length.
    Figure from Ref.~\cite{Frankfurt:1997ss}.}
	\label{fig:rhoDiffCS}
\end{figure}

In Fig.~\ref{fig:rhoDiffCS}~\cite{Frankfurt:1997ss}, the $m=0$ projection is used to calculate the differential cross section for exclusive $\rho$ photoproduction as a function of  $-t$ for almost the full range of photon energies available in Hall D.  
The dotted curve is the Born calculation, which corresponds to scattering from a single nucleon.
For values around $t\approx -0.5$\,GeV$^2$ it displays a strong dependence on the energy of the incident photon due to the latter's coherence length.
The solid curve is the complete vector meson dominance including double-scattering.
It can be seen that introducing double-scattering  generally increases the high $|t|$ cross section, moves the dip region lower in $|t|$ and increases the $\nu$ dependence in the dip region in this energy range.
Double-scattering gives the cross section dip region a strong $\nu$ dependence because the longitudinal interaction length increases with $\nu$.
Studying the $\nu$ dependence of this dip gives direct access to the behavior of the coherence length as a function of energy.  The position of the dip in $-t$ can also be extracted and is related to the diffraction minimum in the charge form factor~\cite{Bosted:1989hy}.

In general, at least three measurements with differing values of both vector and tensor polarization are needed to extract the contributions from the $m=-1,0,+1$ projections independently. 
In this case we expect to run four configurations, two with positive tensor polarization but with opposite vector polarizations, and two with negative tensor polarization, again with opposite vector polarizations.  This will automatically cancel any vector polarization asymmetries.
It would be ideal to alternate between fully populated and fully depleted $m=0$ states, which correspond to tensor polarization values $Q$ of -2 and 1, respectively. 
Unfortunately, standard techniques for solid polarized targets cannot produce $Q<0$, and the practically achievable populations are limited to values between $N_0=0.33$ ($Q=0$) and $N_0=0.17$ ($Q=0.5$).  
However, adiabatic fast passage can transfer $\pm1$ spins into the $m=0$ state and increase its population to about 0.57 ($Q=-0.7$).  Doing so increases the lever arm of the tensor asymmetry by a factor of 2.5, and reduces the beam time required to achieve a specified statistical uncertainly by more than a factor of six.

For projecting to the $m=0$ state at $Q=-2$, this will also decrease the projection distance by a factor of 1.5 over an extraction without negative tensor polarization.
Other final states are likely to benefit similarly.
More importantly, and particularly for the case of $\rho$ mesons, this also decreases the impact of systematic uncertainties related to polarimetry and the presence of backgrounds.

Photoproduction of $\rho$ mesons in this energy range represents about 10\% of the hadronic cross section.  Hence, this reaction can be studied with high statistical precision using a few months of beam time in a single measurement campaign---Jefferson Lab typically has about 25 weeks of beam available per year.  Similar studies with $\omega$ and $\phi$ mesons are also possible, although with lower statistics.

\section{Negative Tensor Polarization and the Deuteron NMR Line Shape}
\label{sec:LineShape}

In spin-1 systems where the three spin projections are populated according to a Boltzmann distribution, the following relation exists between the two polarizations:
\begin{equation}
\label{eq:PzzfromPz}
    Q = 2 - \sqrt{4 - 3P^2}.
\end{equation}
This relation will hold true whenever an ensemble of spin-1 nuclei is in thermal equilibrium with the lattice at a uniform temperature $T$.  It has also been found to be true in many solid systems in which the vector polarization has been enhanced using Dynamic Nuclear Polarization (DNP).  In fact, DNP is frequently described as the cooling of a spin system to a ``spin temperature'' $T_s$ (positive or negative) far below the lattice temperature.  In these circumstances, Eq.~\ref{eq:PzzfromPz} implies that a negative tensor polarization cannot be obtained from dynamic polarization alone and limits the maximum $m=0$ population to that of an unpolarized ensemble,  $N_0 = 1/3$.

One technique to significantly alter $Q$ from this ``Boltzmann restriction'' is partial RF saturation of the nuclear magnetic resonance (NMR) line~\cite{Meyer:1984, Delheij:1996, Keller:2017dll}.   Here, RF is used at strategic locations on the line to equalize the $N_{0}$ population with either $N_{+}$ or $N_{-}$, thereby increasing or decreasing $Q$ from its initial value given by Eq.~\ref{eq:PzzfromPz}.  While this technique can drive the tensor polarization towards zero and can be used simultaneously with dynamic polarization, it cannot produce a significant degree of negative tensor polarization.

A second technique is RF spin transfer from a reservoir of polarized protons.  As described by de~Boer {\em et al.\/} \cite{deBoer:1973}, an enhanced degree of deuteron tensor polarization, positive or negative, can be generated when a mixture of deuterons and protons is dynamically polarized in the conventional manner, and the sample is then irradiated with RF near the proton resonance frequency. The enhanced tensor polarization is found to relax towards zero with an exponential time constant that is strongly temperature dependent, reaching 5.5 hours at 0.1 K and 2.5 T.  Similar results were reported by Delheij {\em et al.\/}~\cite{Delheij:1996}.  However, this technique requires a significant concentration of free, polarizable protons in the target sample.   

A third technique for altering the tensor polarization
of a deuteron target is Adiabatic Fast Passage.  Unlike
partial RF saturation, AFP can produce high degrees of positive or negative $Q$, and unlike RF spin transfer does not require a concentration of protons within the sample.  To describe this method and how it can enhance the tensor polarization, we first discuss the NMR line shape of quadrupolar-broadened deuterons.

In a static magnetic field $\mathbf{B_0}$, the Zeeman interaction will produce three states with energies $E_D = -mh\nu_D$. Here $\nu_D =\mu B_0/h $ is the deuteron Larmor frequency, and $\mu$ is its magnetic dipole moment. RF-induced transitions between the $m=+1$ and 0 states will occur at the same frequency as those between the $m=-1$ and 0 states, and a single resonance line will be observed.  However, the deuteron also possesses an electric quadrupole moment $eQ$, and this will couple to electric field gradients within the sample to produce an additional interaction with energy
\begin{equation}
	E_Q = h\nu_Q[(3 \cos^2\theta - 1) + \eta \sin^2\theta\cos2\phi](3m^2 - 2)   \label{eqn:QuadEnergy}
\end{equation}
where $\nu_Q = \frac{1}{8}\frac{eQeq}{h}$ is the quadrupole frequency, $eq$ is the magnitude of the electric field gradient, and $\theta$ is the polar angle between the gradient and the static magnetic field.  For deuterons the gradient can be assumed to lie along the direction of molecular bonds, for example the ND bond in ammonia and the CD and OD bonds in butanol~\cite{Grutzner}. The azimuthal angle $\phi$ and parameter $\eta$ are necessary when the gradient is not symmetric about the bond.  

The resulting energy levels, assuming azimuthal symmetry, are shown in Fig.~\ref{fig:Levels} and depend both on the magnetic quantum number $m$ and the angle $\theta$.   For a given value of $\theta$, the frequency required to drive the $0 \leftrightarrow \pm 1$ transitions varies as
\begin{equation}
    \nu_\epsilon(\theta) = \nu_D + 3 \epsilon \nu_Q [1 - 3 \cos^2\theta],
\end{equation}
where $\epsilon = \pm 1$ labels the two possible transitions.  The intensity of the transition varies in frequency according to the probability of finding a deuteron with the corresponding polar angle $\theta$.  In amorphous or polycrystalline solids, this angle may take any orientation, with a probability proportional to $1/\cos\theta$.  The result is two overlapping lines that take the form of a Pake doublet~\cite{Pake_1948}, with peaks corresponding to a bond angle $\theta = \pi/2$ and pedestals $\theta = 0$.  In crystals with cubic symmetry (e.g. LiD or HD), $E_Q$ averages to zero and only a single line is observed.  

A more exact expression for the transition lineshape which includes asymmetric bonds and dipolar broadening due to neighboring nuclear and electronic spins is presented by Dulya~{\em et al.}\cite{SpinMuon:1997jxq}:
\begin{align}
\begin{split}
\label{eq:shape}
f_\epsilon(R)=&\frac{1}{2\pi X}\Biggl[ 2 \text{cos}\left(\frac{\alpha}{2}\right) \Biggl(\text{arctan}\Biggl(\frac{Y^2-X^2}{2YX\text{sin}(\frac{\alpha}{2})}\Biggr)+ \frac{\pi}{2}\Biggr)\\
& +\text{sin}\left(\frac{\alpha}{2}\right) \text{ln} \Biggl( \frac{Y^2-X^2+2YX\text{cos}(\frac{\alpha}{2})}{Y^2-X^2-2YX\text{cos}(\frac{\alpha}{2})} \Biggr)\Biggr] 
\end{split}
\end{align}
with 
\begin{align}
\begin{split}
X^2 & = \sqrt{\Gamma^2 + (1 - \epsilon R - \eta\text{cos}(2\phi))^2} \\
Y & = \sqrt{3-\eta\text{cos}(2\phi)} \\
\text{cos}\alpha & = \frac{1 - \epsilon R - \eta\text{cos}(2\phi)}{X^2}
\end{split}
\end{align}
where $R = (\nu - \nu_D)/3\nu_Q$ is the normalized frequency shift, and $\Gamma$ is the width of the Lorentzian function used to describe dipolar broadening. Note that $f_\pm(R) = f_\mp(-R)$.

\begin{figure}[tb]
    \includegraphics[width=0.5\textwidth]{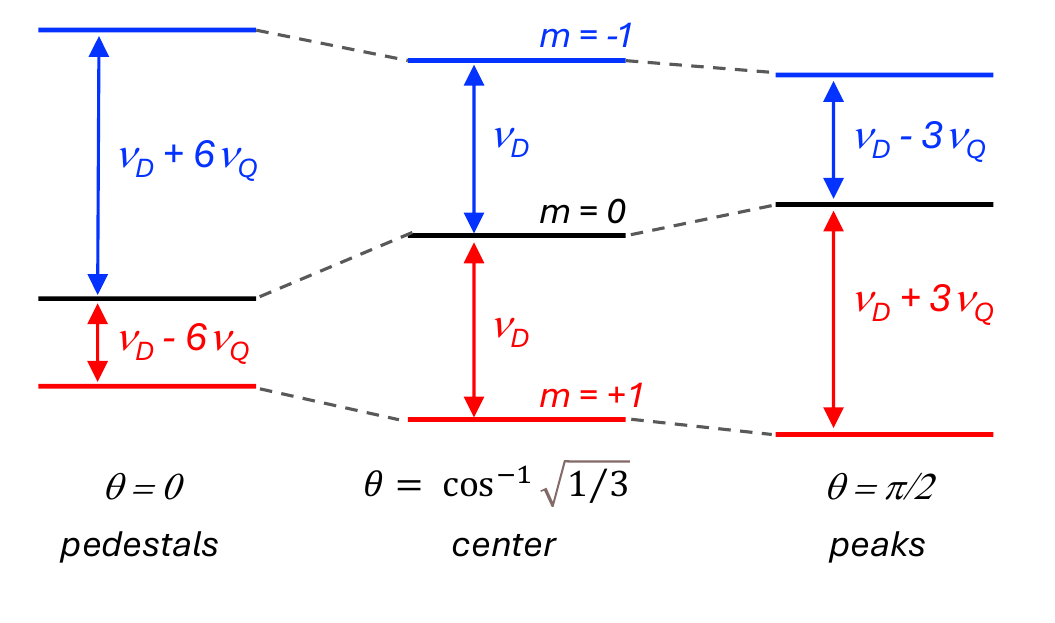}\\
    \includegraphics[width=0.5\textwidth]{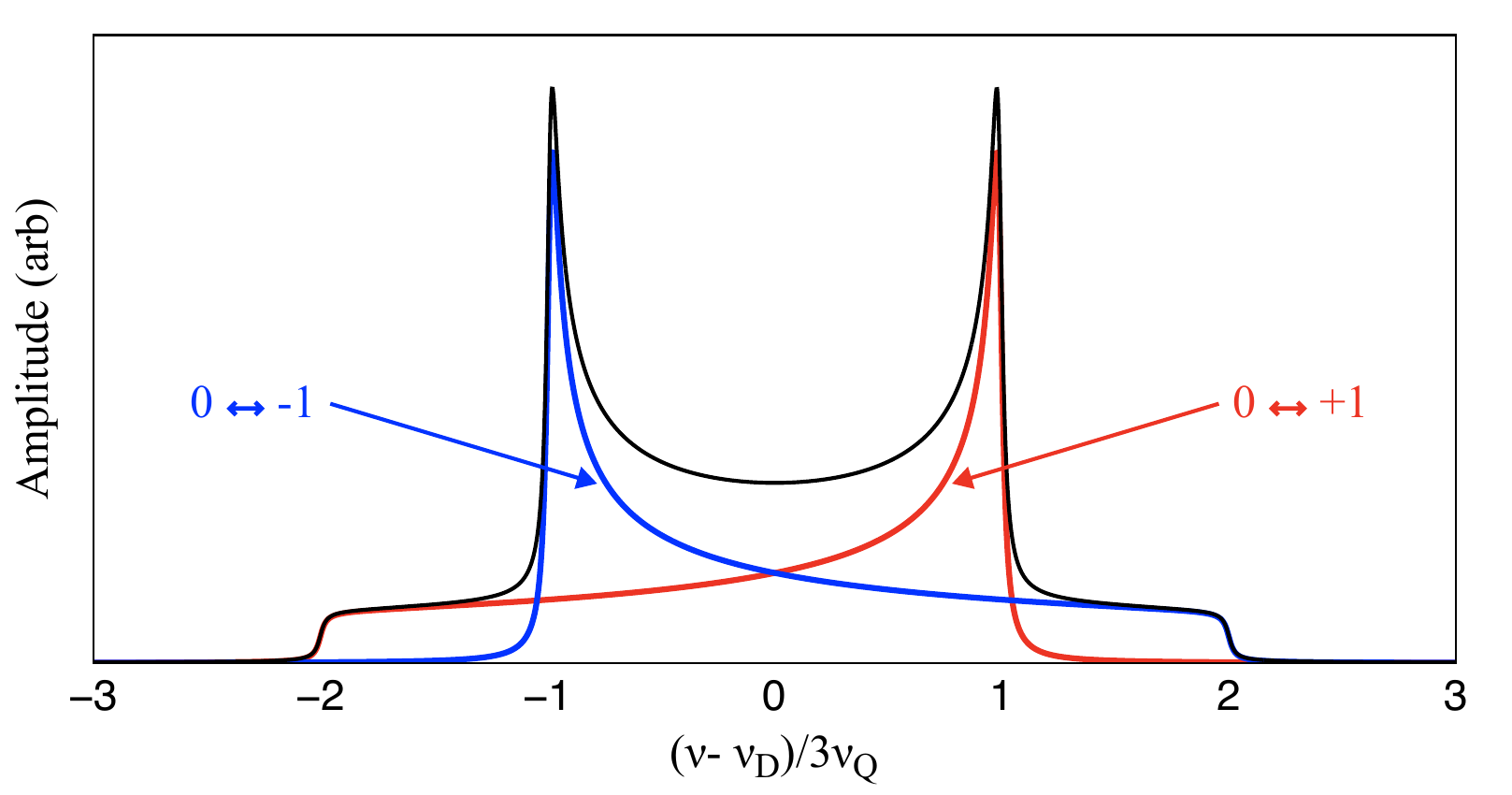}\\
	\caption{Top: Energy level diagram of the deuteron spin system for three values of $\theta$. Bottom: The NMR lineshape, plotted as a function of the frequency shift $(\nu - \nu_D)/3\nu_Q$, is the sum of the $m=0 \leftrightarrow -1$ (blue) and $m=0 \leftrightarrow +1$ transitions.}
	\label{fig:Levels}
\end{figure}

The observed NMR spectra is then the sum of two line shapes, each weighted by the difference in the populations that take part in the transition,
\begin{eqnarray}
    I_{+}(R) & = & c f_{+} (R) [N_{+}(R) - N_{0}(R)] \label{eq:I+}\\
    I_{-}(R) & = & c f_{-} (R) [N_{0}(R) - N_{-}(R)] \label{eq:I-}.
\end{eqnarray}
The vector and tensor polarizations of deuterons with a specified polar angle are given by the sum and difference of $I_{+}(R)$ and $I_{-}(-R)$, respectively, where $R$ is evaluated at $\epsilon (1-3 \cos^2\theta)$.  For example, the vector polarization of deuterons with $\theta = \pi/2$ is $c[I_{+}(1) + I_{-}(-1)]$.  The average polarization of all deuteron spins is 
\begin{eqnarray}
\label{eq:PzIntegral}
    P & = & \int^{+\infty}_{-\infty} [I_{+}(R) + I_{-}(R)]dR \\
    \label{eq:PzzIntegral}
    Q & = & \int^{+\infty}_{-\infty} [I_{+}(R) - I_{-}(R)]dR
\end{eqnarray}
The constant of proportionality $c$ may be determined experimentally by measuring the NMR signal when the sample is polarized under known, thermal equilibrium conditions.

Relations \ref{eq:PzIntegral} and \ref{eq:PzzIntegral} hold irrespective of whether the magnetic levels
are populated according to a Boltzmann distribution or not.
The first of these is the familiar result that the vector polarization is proportional to the total area under the NMR absorption line, which can be directly measured.  The tensor
polarization cannot be measured directly from the NMR line and must be extracted by fitting the line according to Eq.~\ref{eq:shape} or assuming that relation \ref{eq:PzzfromPz} holds.

\begin{figure*}[t]
    \includegraphics[width=1.0\textwidth]{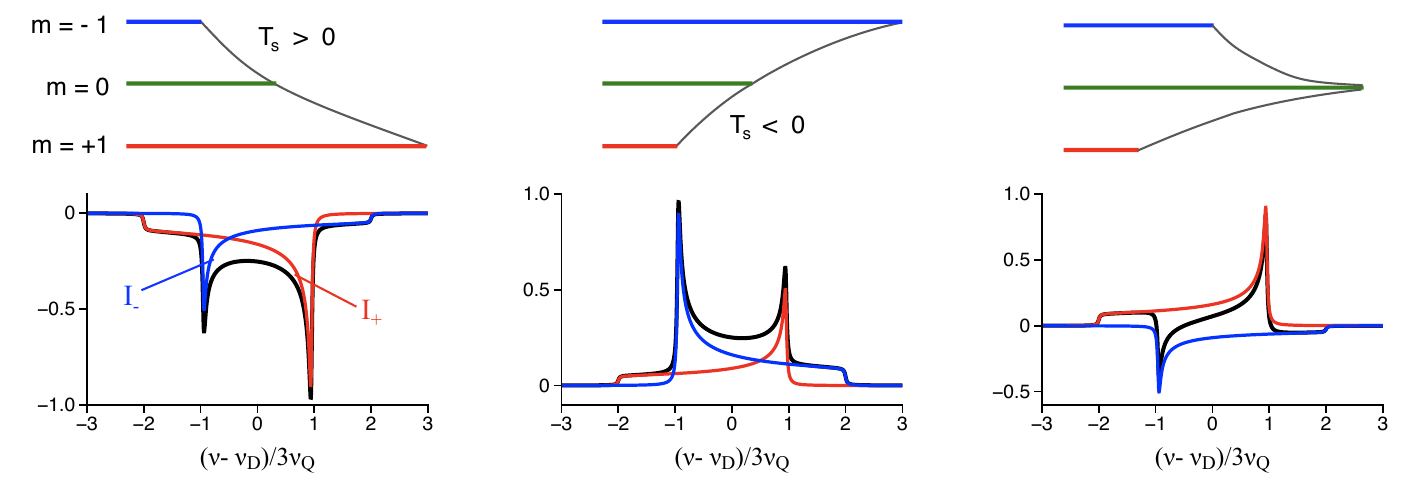}
	\caption{Three representative distributions of deuteron spin systems and their corresponding NMR spectra. From left: positive vector polarization, negative vector polarization, and negative tensor polarization.  In each case, the observed NMR signal (black curve) is the sum of the $I_{+}$ (red) and $I_{-}$ (blue) transitions.  See text for details.}
	\label{fig:ThreeNMR}
\end{figure*}
Representative spectra for three examples are shown in Fig.~\ref{fig:ThreeNMR}.  In the first, the level populations are described by a Boltzmann distribution with a positive spin temperature $T_s>0$, meaning $(N_{+} > N_{0} > N_{-})$. The vector polarization is positive, and the NMR line displays a net absorption of RF energy. In the second, the spin temperature is negative, and $(N_{+} < N_{0} < N_{-})$.  The vector polarization is negative, and there is net emission of RF energy.  In both cases, the tensor polarization is positive.  In the third example, $(N_{0} > N_{-} > N_{+})$, and both vector and tensor polarizations are negative.  This distribution cannot be described by an unique spin temperature, and the resultant NMR signal is a mixture of both absorptive and emissive line shapes.  As we will see below, a deuteron line shape with the same attributes and high degree of negative tensor polarization can be obtained using a standard NMR technique known as Adiabatic Fast Passage.

\section{Adiabatic Fast Passage}
\label{sec:AFP}
Adiabatic Fast Passage (AFP) is a powerful tool to efficiently exchange the populations of adjacent quantum states. In magnetic resonance, AFP is usually performed by applying an RF field slightly above or below the spin resonance frequency and sweeping the static field through resonance.  Alternatively, the field can be fixed, and the RF swept through resonance.  For efficient transfer of one Zeeman state to the next, the adiabatic condition
\begin{equation}
    \label{eqn:Adiabatic}
    \frac{dH_o}{dt} \ll \gamma H^2_1
\end{equation}
should be met, where $dH_0/dt$ is the sweep rate of the static field, $H_1$ is the component of the RF field perpendicular to $H_0$, and $\gamma$ is the gyromagnetic ratio of the nuclear species.  Also, the sweep rate should be faster than the nuclei's
spin-lattice relaxation time $T_1$.  In the case of spin-1/2 nuclei, a single sweep through resonance interchanges the populations of the two Zeeman levels and reverses the polarization.  

For spin-1 deuterons, both the $m=+1$ and $-1$ populations are interchanged with the $m=0$ population, and in quadrupolar broadened systems the result is more complex. As an example, we sweep the RF through the NMR line of a deuteron spin system with from low to high frequency and examine the behavior of the spins with $\theta \sim 0$.  At the low frequency pedestal (between $R\approx -2$ and $-1$), the $m=+1$ and 0 populations are interchanged.  When the RF reaches the high frequency pedestal, the $m=0$ population is exchanged a second time, with the $m=-1$ population. The result is a two-step process that exchanges the populations of the $(-1, 0 +1)$ levels in a cyclic manner: 
$(N_{-}, N_{0}, N_{+}) \rightarrow (N_{-}, N_{+}, N_{0})
\rightarrow (N_{+}, N_{-}, N_{0})$.

A similar exchange occurs for deuterons with $\theta \sim \pi/2$, but the order is reversed.  At the low frequency peak, the $m=-1$ and 0 populations
are exchanged, and the $m=0$ and $m=+1$ populations are exchanged when the RF reaches the high frequency peak:
$(N_{-}, N_{0}, N_{+}) \rightarrow (N_{0}, N_{-}, N_{+})
\rightarrow (N_{0}, N_{+}, N_{-})$.  In the overlap region between the peaks, both exchanges occur, but never for the same value of $\theta$ simultaneously.  An exception is the center of the NMR line, where the three populations will be driven toward equality and the line saturated.

Hautle~{\em et al.\/} have demonstrated the adiabatic fast passsage mechanism in samples of fully-deuterated butanol dynamically polarized at 2.5~T and 0.5~K~\cite{Hautle:1992}.  The results are shown in Fig.~\ref{fig:HautleSweeps}.  Starting with a positive vector polarization of approximately 0.25 (a), RF is swept through resonance from low to high frequency.  The resulting NMR signal (b) is only partially inverted.  The low-frequency portion of the line displays positive vector polarization, while the high-frequency half is negative.  Likewise, the tensor polarization is mixed: $Q$ is positive for the pedestal regions and negative for the rest of the signal (compare to the rightmost line shape in Fig.~\ref{fig:ThreeNMR}).  After an additional RF sweep from low frequency to the line center, the signal is fully inverted, and the vector polarization is reversed.

\begin{figure}
    \centering
    \includegraphics[width=0.35\textwidth]{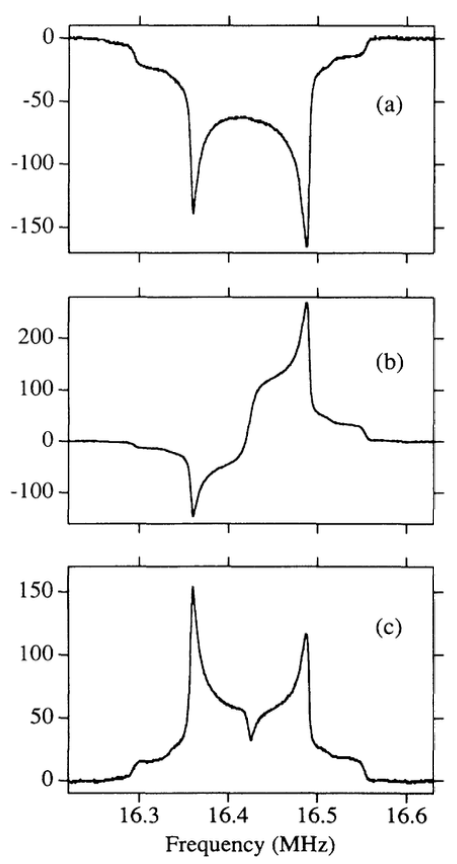}
    \caption{The AFP process in d$_{10}$-butanol.  Figure reprinted from \cite{Hautle:1992}.  See text for details.}
    \label{fig:HautleSweeps}
\end{figure}

In the following section, we model the AFP process in a quadrupolar broadened system of deuteron spins and demonstrate that a high degree of negative tensor polarization can be achieved with a partial AFP sweep.

\section{Modeling Adiabatic Fast Passage}
\label{sec:Model}

In order to model the AFP process we specify the distribution of the three magnetic substates as a function of the reduced frequency $R$.  We assume the states are populated according to a Boltzmann distribution at a common spin temperature, such that Eq.~\ref{eq:PzzfromPz} is valid.  The fractional populations can then be calculated according to Eq.~\ref{eq:N+}--\ref{eq:N-}.  The distribution of the $N_i$ spins that take part in the $0 \leftrightarrow \pm1$ exchange is then
\begin{equation}
    n_i(R) = N_if_{\pm}.
\end{equation}
The population distributions for the $0 \leftrightarrow \pm1$ exchanges, assuming a polarization $P = 0.75$, is shown in Fig.~\ref{fig:Pz75_states}.

This shape is then used to model the number of nuclei in each spin projection as a function of R, with $n_{+}(R)$, $n_{-}(R)$, and $n_{0}(R)$, the number of states in  $m=1,-1$ and $0$ respectively. 
For the purposes of the model we choose to think of the spins located in frequency space rather than as a function of the angle $\theta$ between the magnetic field and the electric field gradient.  Figure~\ref{fig:Pz75_states} shows these states as a function of a dimensionless position in the NMR line,  spanning the domain of the NMR signal, $R = (\omega - \omega_D)/3\omega_Q$.

\begin{figure}[htb]
	\includegraphics[width=0.40\textwidth]{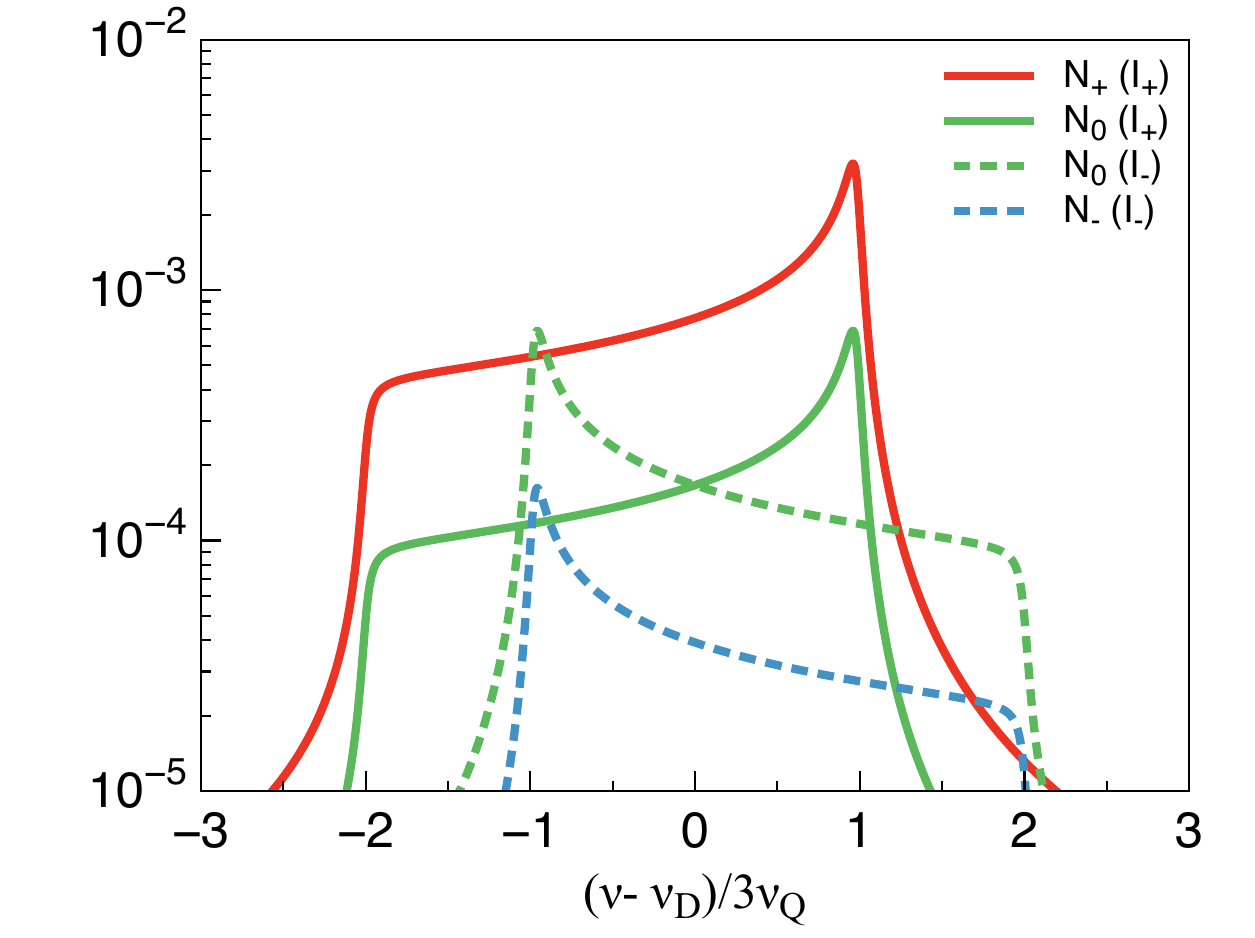}
	\caption{The density of states as a function of the reduced frequency $R$, for a vector polarization $P=0.75$.  The states taking part in the $I_+$ ($0 \leftrightarrow +1$) are shown with solid lines, while those for the  $I_-$ ($0 \leftrightarrow -1$) transition are shown with dashed lines.  Note the logarithmic scale.   }
	\label{fig:Pz75_states}
\end{figure}

From the model of the underlying distribution of states, one can reconstruct the NMR lineshape from $I_{+}(R) = n_{0}(R,+1) - n_{+}(R,+1)$ and $I_{-}(R) = n_{-}(R,-1) - n_{0}(R,-1)$.  The resulting distributions can be seen in Fig.~\ref{fig:Pz75_pake}.  Here we assume a conservative deuteron vector polarization of $P=0.75$, the proportion of nuclei in three Zeeman states will be $N_{+}=0.79$, $N_{-}=0.04$, and $N_{0}=0.17$, which gives a tensor polarization $Q = 0.48$.

\begin{figure}[htb]
	\includegraphics[width=0.40\textwidth]{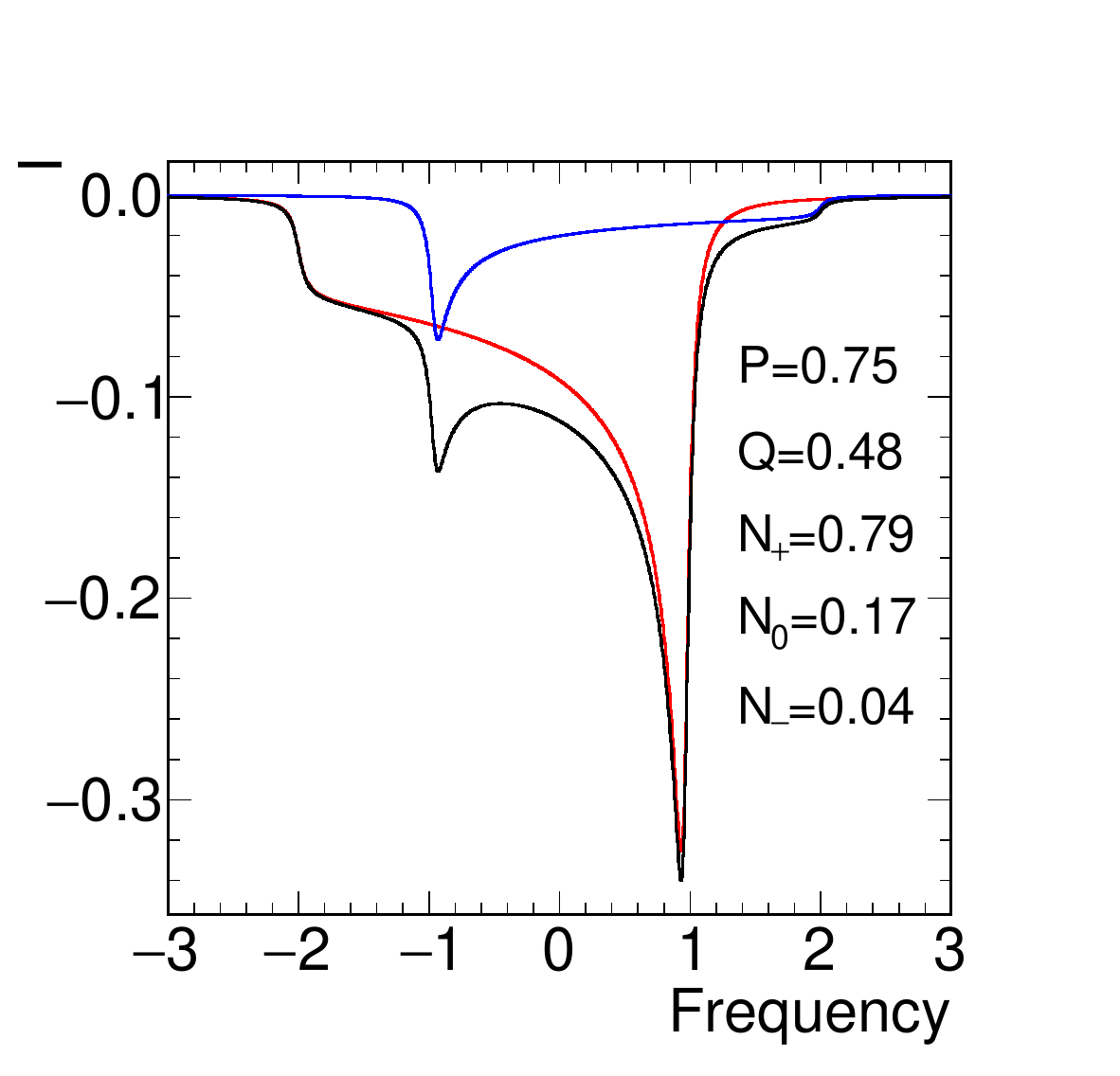}
	\caption{The distributions $I_{+}$, $I_{-}$ and the sum as function of R for $P=0.75$.}
	\label{fig:Pz75_pake}
\end{figure}

We can use this model to apply various transformations to predict what the resulting values of the tensor polarization is.
We verified that we reproduce the qualitative features of data from Ref.~\cite{Hautle:1992}.
Figure~\ref{fig:Hautle} shows NMR lineshapes for 3 configurations, panel (a) an initial state polarized to $P=0.25$.  Panel (b) shows the lineshape for after a single AFP sweep through the full frequency range from low to high which generates a negative tensor polarization.  Panel (c) shows another half sweep from from low to high.
The most obvious feature that is not reproduced is the area of the central frequency in (b) and (c) where there appears to be a region of depolarization.  This likely arises from simultaneously driving the $-1\leftrightarrow0$ and $0\leftrightarrow+1$ transitions at the same value of $\theta$ which tends to drive the populations to equilibrium leading to depolarization.
In progressing from (a) to (c), Ref.~\cite{Hautle:1992} found a 0.91 efficiency in transfer of polarization from positive vector to negative vector.

\begin{figure}[htb]
\begin{subfigure}{.25\textwidth}
  \includegraphics[width=.99\linewidth]{Hautle_sweeps.png}
\end{subfigure}%
\begin{subfigure}{.25\textwidth}
  \includegraphics[width=.99\linewidth]{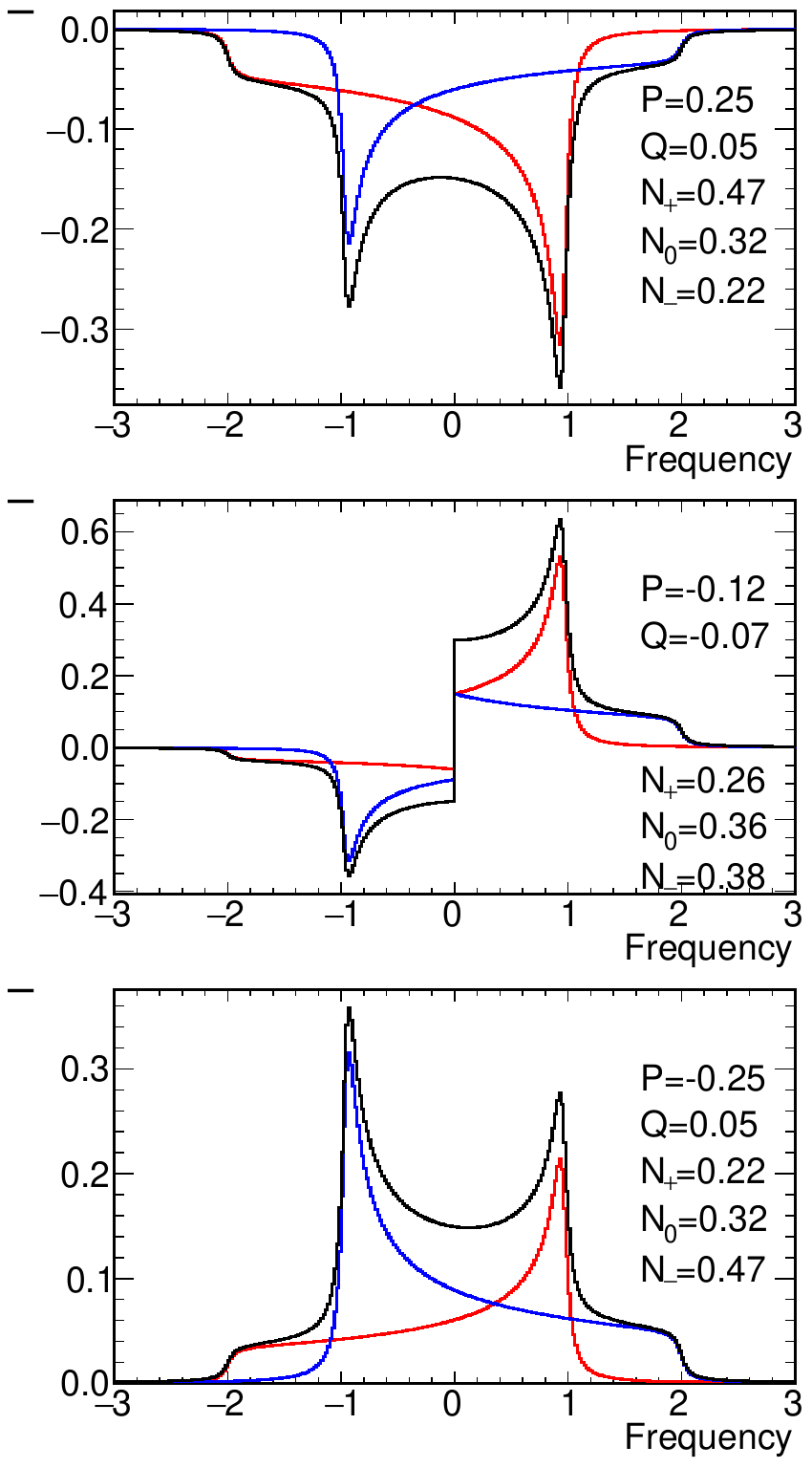}
\end{subfigure}
\caption{A comparison of data from Ref.~\cite{Hautle:1992} (left) with the model (right).  (a) Initial signal with $P=0.25$.  (b) After a full sweep from low to high frequency. (c) After an additional half-sweep from the low to the center frequency.  This demonstrates that the model reproduces the general features of AFP.}
\label{fig:Hautle}
\end{figure}

The success of the model allows the determination of a strategy to maximize negative tensor polarization.
For high initial polarization a simple strategy is likely to be optimal.  
For $P=0.75$, performing an incomplete AFP scan from low to high frequency, stopping at $R=1.04$ produces $Q=-0.71$ with $N_{0}=0.57$, see Fig.~\ref{fig:Pz75_AFP_pake}.  The corresponding populations of the spin states as a function of frequency can be seen in Fig~\ref{fig:Pz75_AFP_states}.

\begin{figure}[htb]
	\includegraphics[width=0.40\textwidth]{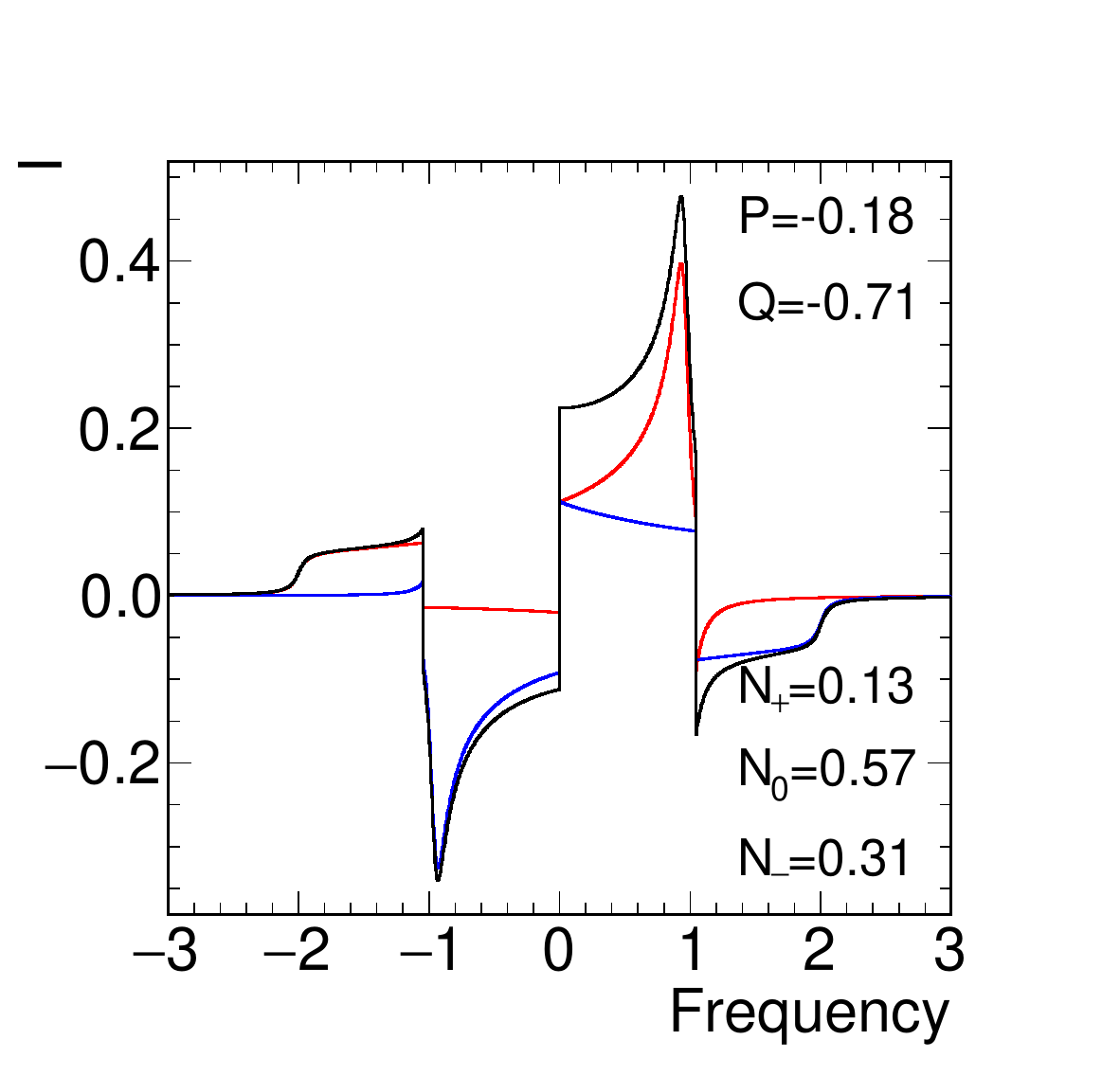}
	\caption{Predicted NMR lineshape for an initial $P=0.75$ followed an AFP scan from low frequency to $R=1.04$.}
	\label{fig:Pz75_AFP_pake}
\end{figure}

\begin{figure}[htb]
	\includegraphics[width=0.40\textwidth]{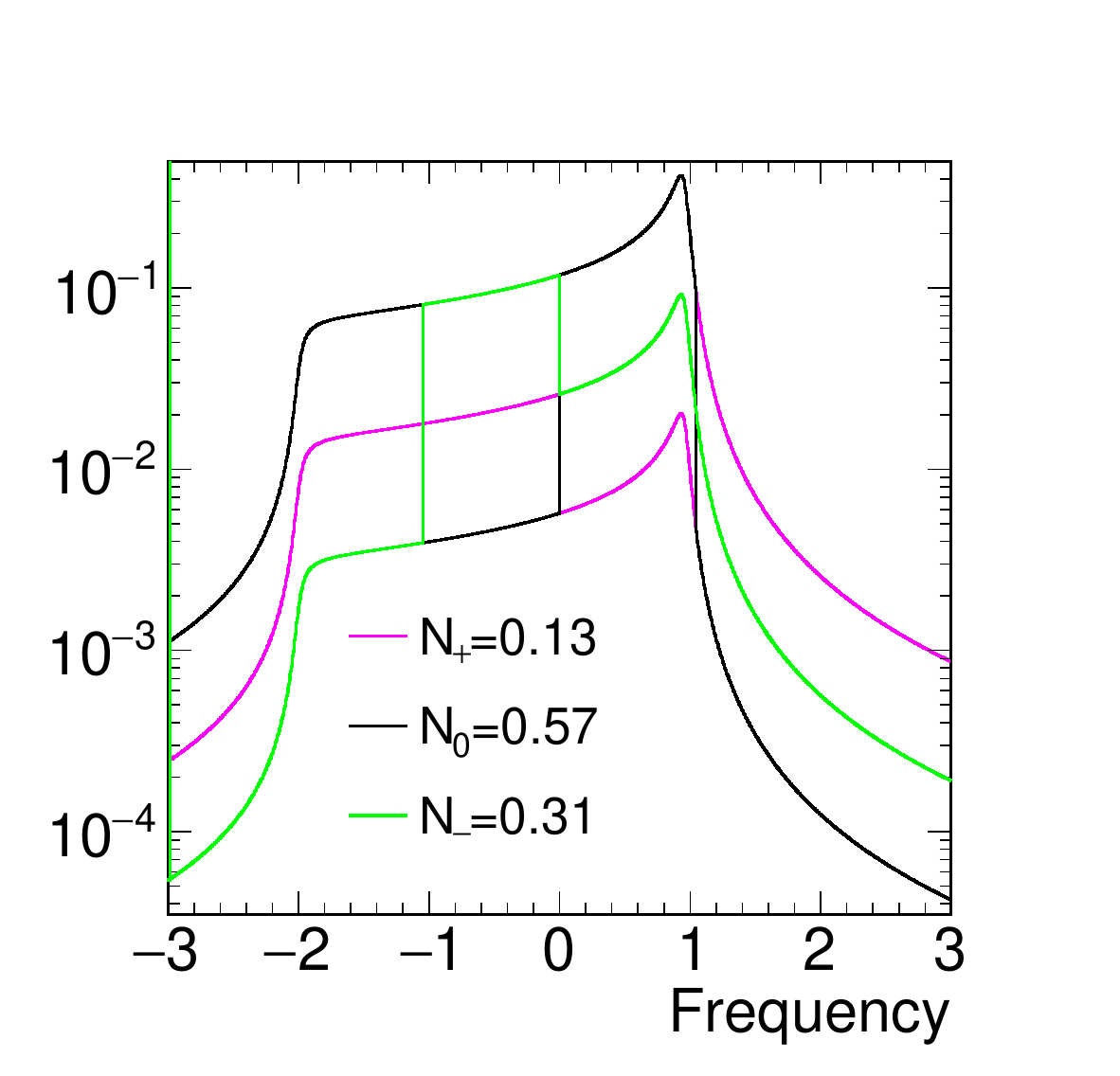}
	\caption{Predicted populations of the spin states for an initial $P=0.75$ followed an AFP scan from low frequency to $R=1.04$. Note the logarithmic scale.}
	\label{fig:Pz75_AFP_states}
\end{figure}

The model is useful for linking observed vector and tensor polarizations to population of the spin state projections and for allowing arbitrary manipulations of the these underlying states.
This model could be improved by modeling the spin state as a function of $\theta$ instead of R and then modeling the interaction with the environment so that the broadening of the spectrum and motion of the peaks are not added empirically.  Such a procedure may be able to account for the features observed in the NMR lineshapes of Fig.~\ref{fig:Hautle}.
This is expected to be important for accurately determining the tensor polarization from the measured sweeps.
Obtaining lineshape data on polarized deuterium before and after AFP manipulations will be necessary to refine the model. 
Likewise, the lifetime of such a non-equilibrium spin state is known to be strongly temperature dependent~\cite{deBoer:1973}.

\section{Target Conceptual Design}
\label{sec:targetdesign}
Dynamically polarized solids can provide a dense, compact source of spin-polarized nuclei such as $^1$H, D, $^6$Li, $^7$Li, and $^{13}$C.  During Jefferson Lab's 6~GeV era, dynamically polarized targets were utilized in experimental halls A, B, and C, and a rich science program will again use these systems at 12~GeV.  This includes Hall D, where the Real Gamma GDH Experiment (REGGE)~\cite{Dalton:2020wdv} has been approved to measure the high-energy behavior of the Gerasimov-Drell-Hearn (GDH) sum rule for photons up to 12 GeV.  
Once it has been constructed for REGGE, the target which is described below can also be used to generate extremely high levels of tensor polarization, both positive and negative, in deuterons.  

Target samples will be polarized in the magnetic field of the Hall D solenoid, with additional correction coils used to increase the field to 2.5~T and decrease its gradient to a level that is acceptable for dynamic nuclear polarization (DNP).  The target system will include a $^3$He-$^4$He dilution refrigerator with sufficient cooling capacity to maintain a sample temperature below 0.3~K with microwaves on during the DNP process and a base temperature below 50~mK to ensure long
spin-relaxation times with the microwaves off.  In this way the sample can be continuously polarized for maximum vector polarization, or the system can operate as a frozen spin target.  As we shall see, the latter operation can provide new opportunities for tensor polarized deuterons.  Under these conditions, vector polarizations up to 80\% have been demonstrated using narrow EPR-line radicals dissolved in glassy, fully deuterated alcohols such as d-butanol~\cite{Goertz:2004}.  Assuming a Boltzmann distribution for the deuteron sublevels, the corresponding tensor polarization is approximately 56\% (Eq.~\ref{eq:PzzfromPz}).  

For optimal dynamic nuclear polarization, one typically aims for a field homogeneity $\Delta B/B$ of about 100~ppm, and assuming a 10 cm long target sample and a field strength of 2.5~T, this implies $dB/dz \sim 25$~$\mu$T/cm.  Unfortunately, the Hall D solenoid does not satisfy this requirement.  The on-axis field of the solenoid and its gradient are shown in Fig.~\ref{fig:Field} with two, 10 cm long regions highlighted.  
The first region, centered at 65~cm, corresponds to the nominal location of the liquid hydrogen target utilized for the GlueX experiment, and here $B\sim1.6~T$ and $dB/dz \sim 5$~mT/cm.  At the second, centered
at $z=130$~cm, $B\sim1.8~T$ and $dB/dz \sim 2$~mT/cm.  At either location,
the field uniformity should be improved by about two orders of magnitude.  We propose to achieve this level of uniformity using a series of thin, superconducting coils inside the target cryostat.

Results of preliminary investigation are shown at the bottom of Fig.~\ref{fig:Field}, where three coils are used to simultaneously increase the field strength to 2.5~T and improve the homogeneity at the 130~cm region to approximately $\pm40$~ppm over the 10~cm length of the target \cite{Lagerquist}.  The design of the coils is detailed in Table~\ref{tab:Shims}.  Successful implementation of this solution will require winding the coils with extreme precision, a practice pioneered by the Bonn Polarized Target Group \cite{BonnCoils}.

\begin{table*}[]
    \centering
    \begin{tabular}{|c|c|c|c|c|c|c|}
    \hline
        Target & Coil &  Coil & Coil & Windings & Number of & Coil\\
        Center (cm) & Number & Center (cm) & Length (cm)  & per Layer & Layers & Current (A) \\
        \hline
        \hline
         130 & 1 & -3.3 &  22.9 & 996  & 4 & 36.3 \\
            & 2 & -1.2 &  14.2 & 617 & 2 & -11.5 \\
            & 3 & 1.1  &  6.4 & 278.3 & 1 & -1.0\\
         \hline    
    \end{tabular}
    \caption{Preliminary winding parameters for three correction coils assuming the target sample is centered 130~cm from the upstream edge of the Hall D solenoid.  Coil center is measured relative to the target's center position.  A negative current produces a magnetic field antiparallel to the solenoid's field.  The wire diameter is 0.23~mm, and the coil diameter is 2.0 cm.}
    \label{tab:Shims}
\end{table*}

\begin{figure}[htb]
	\includegraphics[width=0.45\textwidth]{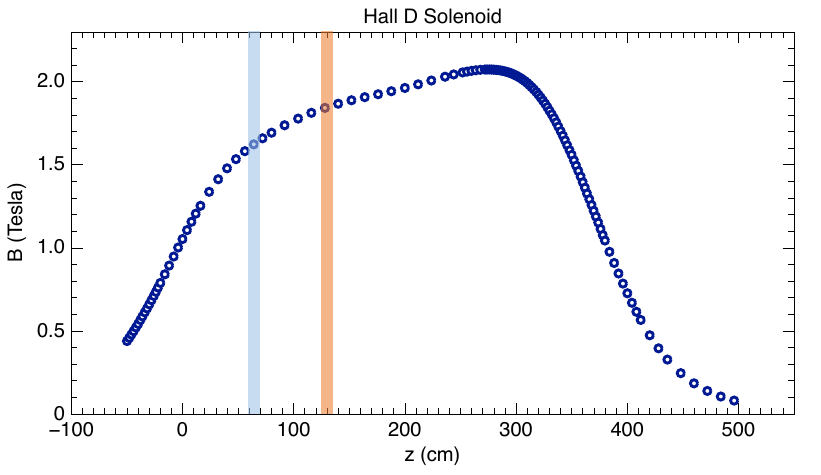}\\
    \includegraphics[width=0.45\textwidth]{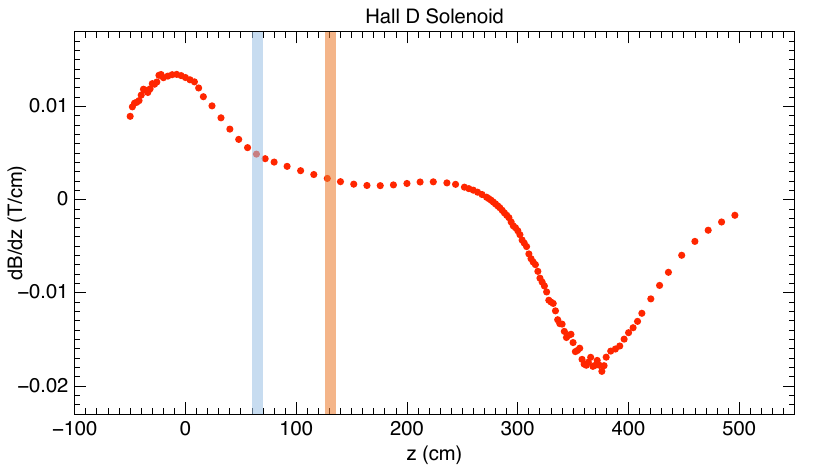}\\
    \includegraphics[width=0.45\textwidth]{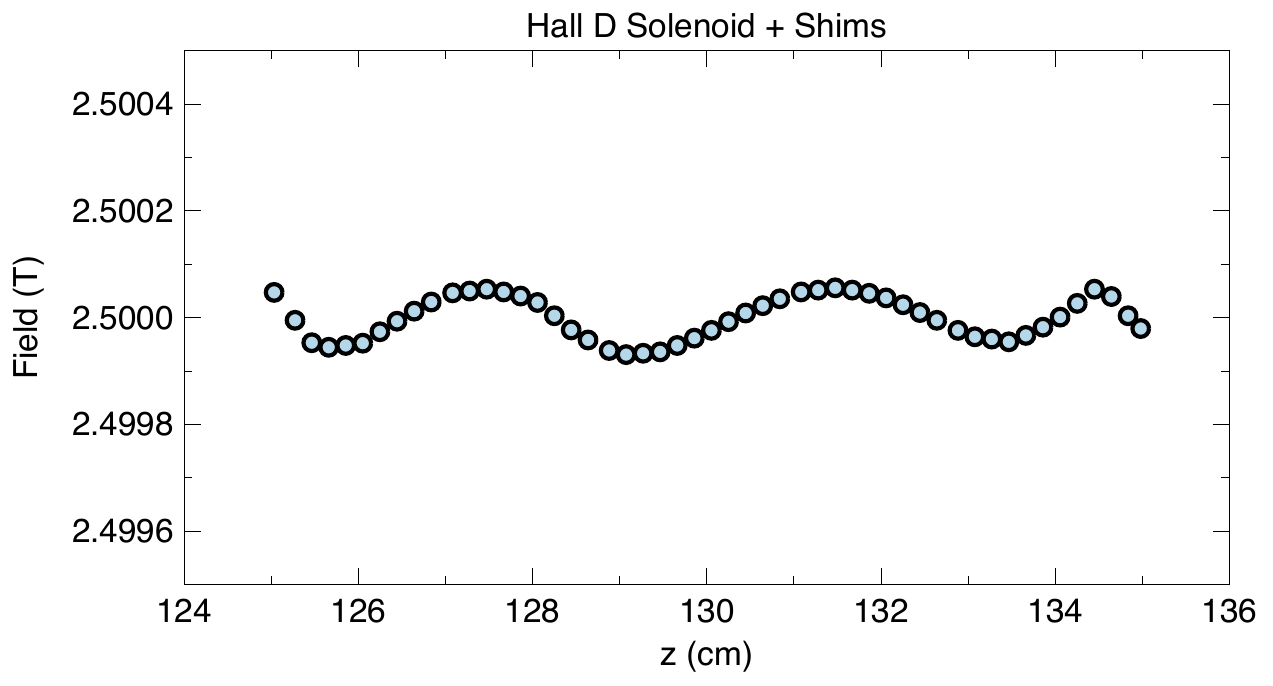}\\
	\caption{Measured, on-axis field strength (top) and field gradient $dB/dx$ (middle) of the Hall D solenoid.  The two highlighted regions are the possible target positions examined in the text.  Calculated field using three shim coils centered at 130~cm.  Distances are measured relative to the upstream edge of the solenoid.}
	\label{fig:Field}
\end{figure}

\section{Summary}

We have discussed the particular case of the coherent photoproduction of the $\rho$ meson from polarized deuterium, for energies from 3 to 12\,GeV, as an example of where deuterium with large negative tensor polarization is both possible and a significant advantage in terms of beam time.
We expect to achieve an improvement of 2.5 on the uncertainties on asymmetries, which represents a factor of 6.2 on the beam time required over using positive tensor polarization alone.
  
The spin-state-separated differential cross section offers a unique laboratory to study double-scattering and the coherence properties of a high energy photon. 
Future work will generalize these ideas to the production of other mesons and to incoherent interactions which would also yield interesting insights, such as measuring the $\phi$-nucleon interaction cross section or high momentum observables related to non-nucleonic degrees of freedom.
In general, studies using real photons will serve as a $Q^2 = 0$ benchmark for future studies on the polarized deuteron using the Electron Ion Collider, see for example Ref.~\cite{Mantysaari:2024xmy}.

Obtaining high deuteron polarization in butanol through DNP and producing negative tensor polarization through AFP have both been demonstrated previously and are well understood.
The use of butanol as a polarized target is mostly limited to photon beams as electrons beam destroy the paramagnetic radical used for polarization.  However, the adiabatic fast passage technique described here is applicable to any quadrupolar-broadened target material, including the more radiation-resistant deuterated ammonia (ND$_3$).  The standard equipment in Hall D, along with a new  polarized target, is perfectly suited to perform these high-impact measurements.

\section*{Acknowledgement}
The authors gratefully acknowledge Victoria Lagerquist (Physikalisches Institut, Bonn), who produced the model of correction coils utilized in this work.   
This material is based upon work supported by the U.S. Department of Energy, Office of Science, Office of Nuclear Physics under contract DE-AC05-06OR23177.

\bibliographystyle{spphys}
\bibliography{bibliography.bib}

\end{document}